\documentclass[twocolumn,journal]{IEEEtran}

\usepackage{amsfonts}
\usepackage{amsmath}
\usepackage{amsthm}
\usepackage{amssymb}
\usepackage{subcaption}

\usepackage{graphicx}
\usepackage[T1]{fontenc}
\usepackage{supertabular}
\usepackage{longtable}
\usepackage[usenames,dvipsnames]{color}
\usepackage{bbm}
\usepackage{fancyhdr}
\usepackage{breqn}

\usepackage{capt-of}
\usepackage{tikz}
\usetikzlibrary{matrix}
\usepackage{endnotes}
\usepackage{soul}
\usepackage{marginnote}
\newcommand{\mathsym}[1]{}
\newcommand{\unicode}[1]{}

\usepackage{colortbl}

\usepackage[framemethod=TikZ]{mdframed}
\usepackage[framemethod=TikZ]{mdframed}
\usepackage{framed}

\mdfsetup{%
	nobreak=true,
	outerlinewidth=1,skipabove=20pt,backgroundcolor=yellow!30, outerlinecolor=black,innertopmargin=5pt,splittopskip=\topskip,skipbelow=\baselineskip, skipabove=\baselineskip,ntheorem,roundcorner=5pt}

\mdtheorem[nobreak=true,outerlinewidth=1,
backgroundcolor=yellow!50, outerlinecolor=black,innertopmargin=0pt,splittopskip=\topskip,skipbelow=\baselineskip, skipabove=\baselineskip,ntheorem,roundcorner=5pt,font=\itshape]{result}{Result}

\mdtheorem[nobreak=true,outerlinewidth=1,
backgroundcolor=yellow!30, outerlinecolor=black,innertopmargin=0pt,splittopskip=\topskip,skipbelow=\baselineskip, skipabove=\baselineskip,ntheorem,roundcorner=5pt,font=\itshape]{theorem}{Theorem}

\mdtheorem[nobreak=true,outerlinewidth=1,
backgroundcolor=gray!10, outerlinecolor=black,innertopmargin=0pt,splittopskip=\topskip,skipbelow=\baselineskip, skipabove=\baselineskip,ntheorem,roundcorner=5pt,font=\itshape]{remark}{Remark}

\mdtheorem[nobreak=true,outerlinewidth=1,
backgroundcolor=gray!10, outerlinecolor=gray!10,innertopmargin=0pt,splittopskip=\topskip,skipbelow=\baselineskip, skipabove=\baselineskip,ntheorem,roundcorner=5pt,font=\itshape]{definition}{Definition}

\mdtheorem[nobreak=true,outerlinewidth=1,
backgroundcolor=pink!30, outerlinecolor=black,innertopmargin=0pt,splittopskip=\topskip,skipbelow=\baselineskip, skipabove=\baselineskip,ntheorem,roundcorner=5pt,font=\itshape]{quaestio}{Quaestio}

\mdtheorem[nobreak=true,outerlinewidth=1,
backgroundcolor=yellow!50, outerlinecolor=black,innertopmargin=5pt,splittopskip=\topskip,skipbelow=\baselineskip, skipabove=\baselineskip,ntheorem,roundcorner=5pt,font=\itshape]{background}{Background}

\mdtheorem[nobreak=true,outerlinewidth=1,
backgroundcolor=gray!10, outerlinecolor=black,innertopmargin=5pt,splittopskip=\topskip,skipbelow=\baselineskip, skipabove=\baselineskip,ntheorem,roundcorner=5pt,font=\itshape]{nothing}{}

\mdtheorem[nobreak=true,outerlinewidth=1,
backgroundcolor=pink!50, outerlinecolor=black,innertopmargin=5pt,splittopskip=\topskip,skipbelow=\baselineskip, skipabove=\baselineskip,ntheorem,roundcorner=5pt,font=\itshape]{point}{Point}
\mdtheorem[nobreak=true,outerlinewidth=1,
backgroundcolor=pink!50, outerlinecolor=black,innertopmargin=5pt,splittopskip=\topskip,skipbelow=\baselineskip, skipabove=\baselineskip,ntheorem,roundcorner=5pt,font=\itshape]{lemma}{Lemma}

\mdtheorem[nobreak=true,outerlinewidth=1,
backgroundcolor=pink!50, outerlinecolor=black,innertopmargin=5pt,splittopskip=\topskip,skipbelow=\baselineskip, skipabove=\baselineskip,ntheorem,roundcorner=5pt,font=\itshape]{commentary}{Comment}

\mdtheorem[nobreak=true,outerlinewidth=1,
backgroundcolor=pink!50, outerlinecolor=black,innertopmargin=5pt,splittopskip=\topskip,skipbelow=\baselineskip, skipabove=\baselineskip,ntheorem,nobreak=true,roundcorner=5pt,font=\itshape]{proposition}{Proposition}

\begin{document}

\title{{\color{Red}
Informational Rescaling of PCA Maps\\
with  Application to Genetic Distance}}
\author{Nassim Nicholas Taleb\IEEEauthorrefmark{1}, Pierre Zalloua\IEEEauthorrefmark{2}\IEEEauthorrefmark{4}, Khaled Elbassioni\IEEEauthorrefmark{2}, Andreas Henschel\IEEEauthorrefmark{2} and Daniel E. Platt\IEEEauthorrefmark{3}\\ \IEEEauthorblockA{\IEEEauthorrefmark{1}  Tandon School, New York University (Corresponding author, nnt1@nyu.edu)\IEEEauthorrefmark{2} Khalifa University \IEEEauthorrefmark{3} Harvard University \IEEEauthorrefmark{4} IBM }
\\
February 2024}

\maketitle
\begin{mdframed}
\bigskip
\begin{abstract}
We discuss the inadequacy of covariances/correlations and other measures in L2 as relative distance metrics under some conditions. We  propose a computationally simple heuristic to transform a map based on standard principal component analysis (PCA) (when the variables are asymptotically Gaussian) into an entropy-based map where distances are based on mutual information (MI). Rescaling Principal Component based distances using MI 
allows a representation of relative statistical associations when, as in genetics,  it is applied on bit measurements between individuals' genomic mutual information. 

This entropy rescaled PCA, while preserving order relationships (along a dimension), changes the relative distances to make them linear to information. We show the effect on the entire world population and some subsamples, which leads to significant differences with the results of current research. 

\end{abstract}
	\end{mdframed}
	
	\begin{figure*}[t!]
\includegraphics[width=\linewidth]{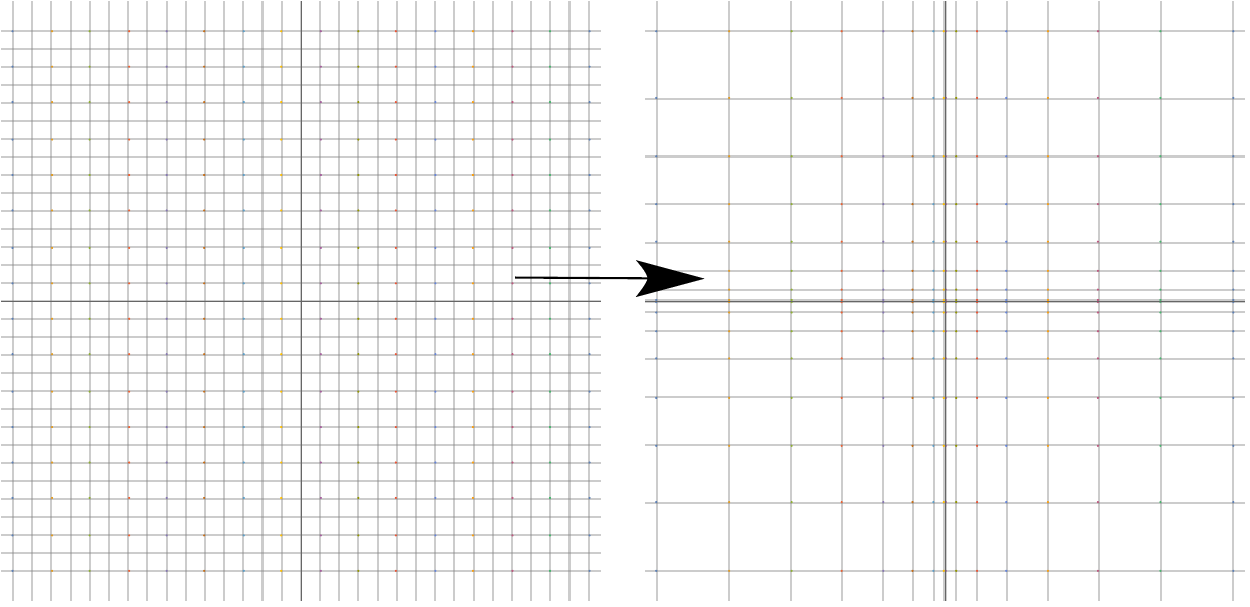}
\caption{Transformation of PCA maps to accommodate informational distances}\label{transformation}
\end{figure*}

\section{Introduction: The problem of correlation}

Correlation between two variables $X$ and $Y$, even if we assumed that both variables are normally distributed (or, more generally, in the class of rapid convergence to the normal, or "thin tailed" thanks to the approximative behavior\cite{taleb2019statistical}), does not adequately reflect the information distance between them. Nor would squared correlation do. This distortion becomes acute with Principal Component Analysis, PCA, and the genetic two-dimensional maps where there is a built-in correlation component.

For instance, if we are correlating 2 vectors $X_1$ and $X_2$ against $Y$ (assuming it is the basis) the information does not scale linearly (even though correlation reflects a measure of the noise in a linear dependence). There must be some scaling of the correlation metric. A $.5$ correlation is vastly --and disproportionally --inferior to, say, $.7 $.


\begin{figure}[h!]
\begin{subfigure}[b]{.46\linewidth}
  \centering
\includegraphics[width=\linewidth]{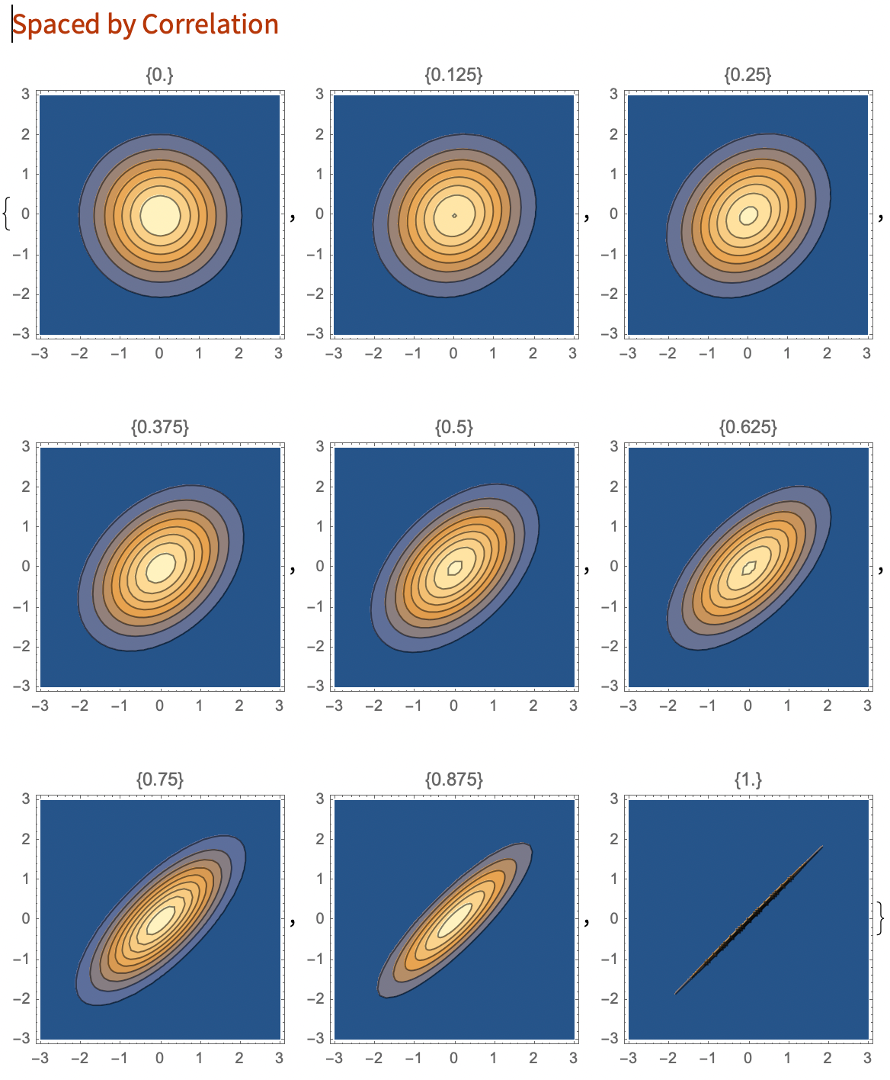}
\label{}
\end{subfigure}
\hfill
\begin{subfigure}[b]{.45\linewidth}
\includegraphics[width=\linewidth]{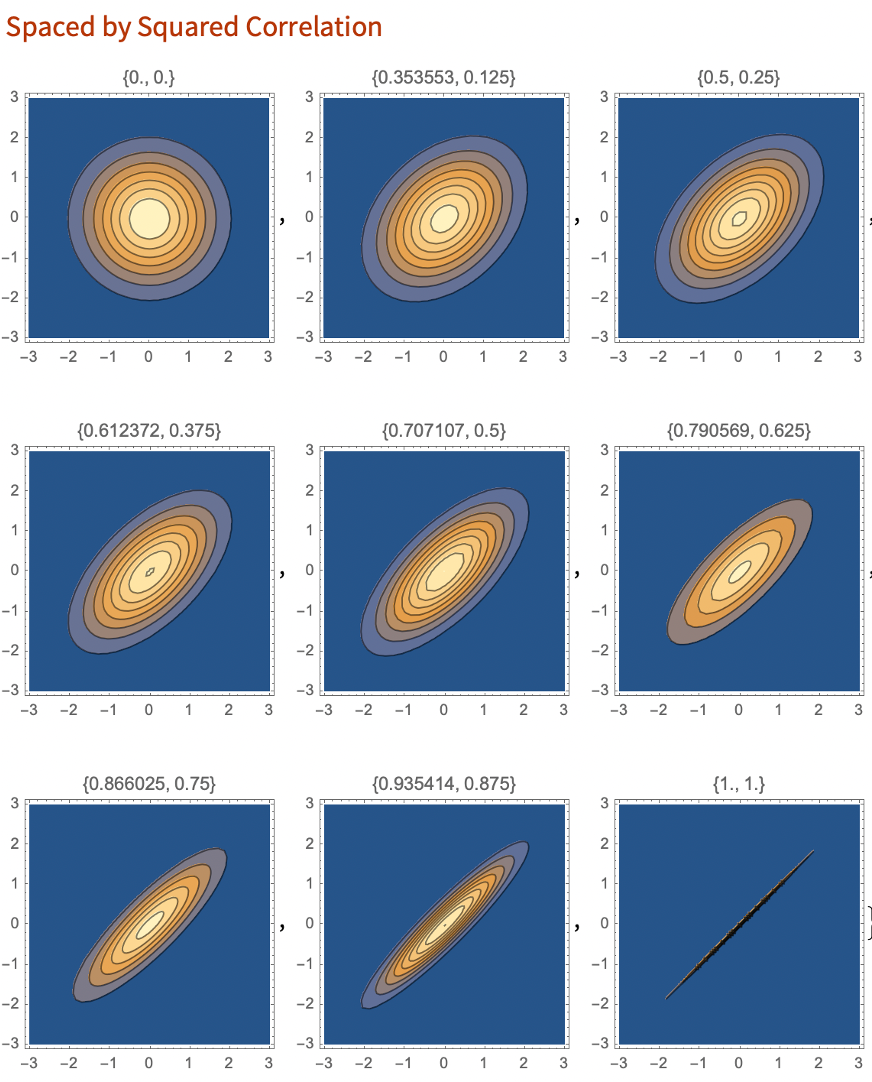}
\label{}
\end{subfigure}
\centering
\begin{subfigure}[b]{.5\linewidth}
\centering
\includegraphics[width=\linewidth]{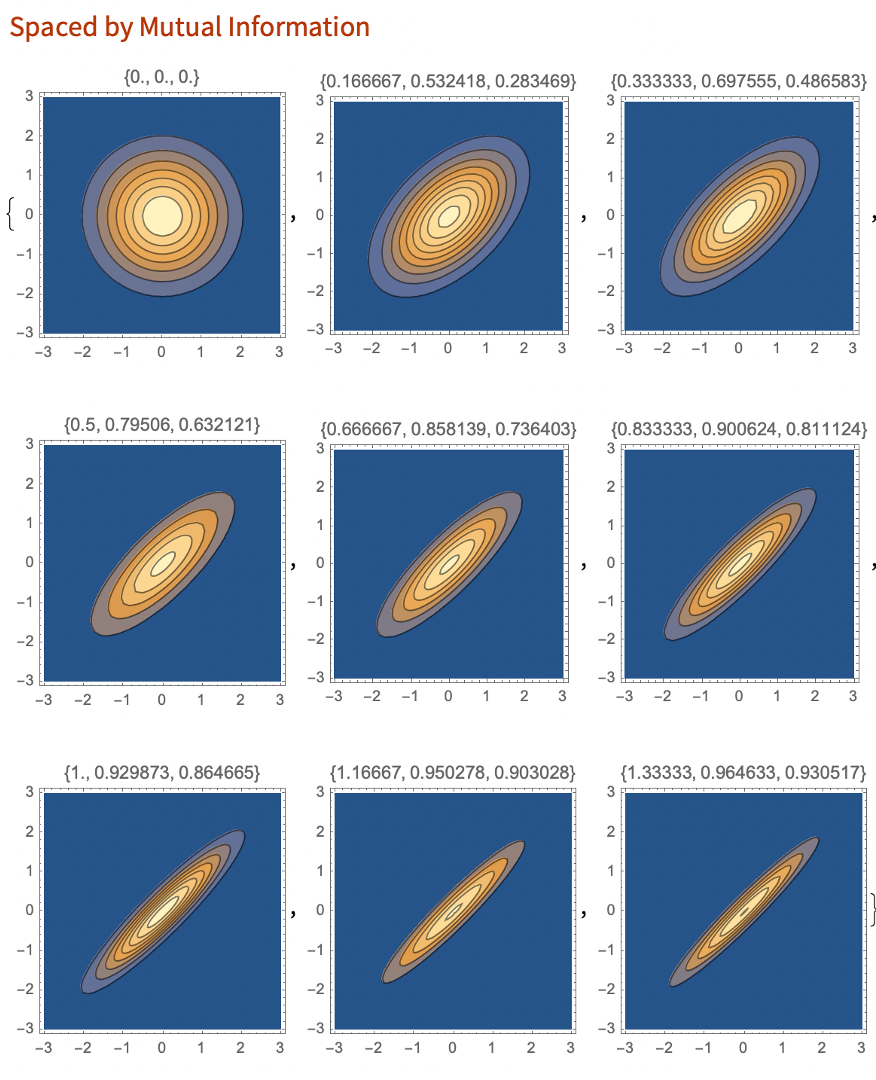}
\label{}
\end{subfigure}
\caption{The visual intuition for the three possible methods for informational distances. We generate bivariate normal distributions for $X$ and $Y$, and represent the iso-densities on the $X$ and $Y$ axes. Each square is equidistant with respect to the parameters correlation, correlation squared, and MI to the one to its left and its right, above and below it, as well as on the diagonal. MI matches our visual intuition.}
\label{fig:PCA-ex2}
\end{figure}


\begin{figure}[!ht]
\centering
\begin{subfigure}[t]{\linewidth}
\includegraphics[width=\linewidth]{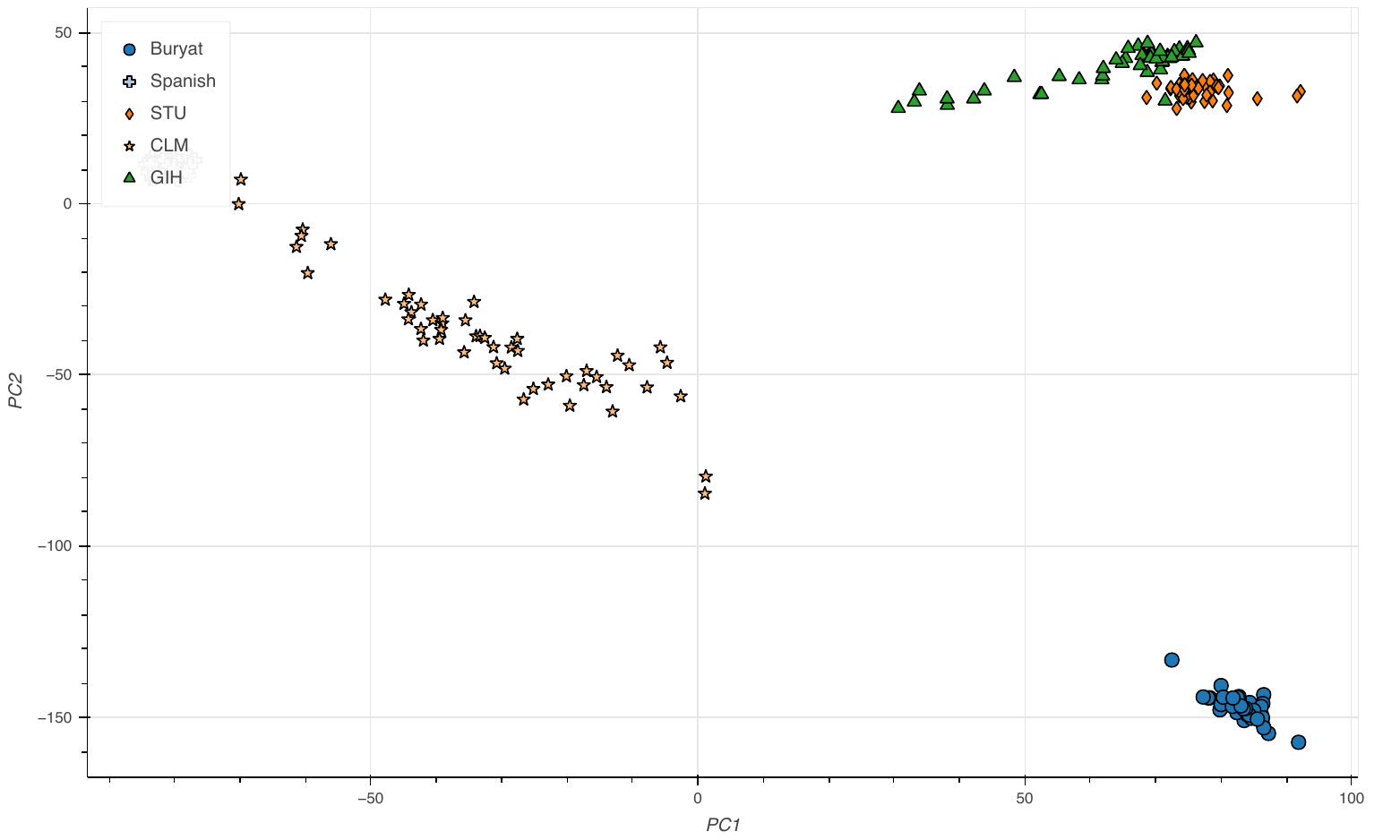}
\subcaption{Correlation-based PCA}
\bigskip
\label{fig:pca12}
\end{subfigure}

\begin{subfigure}[t]{\linewidth}
\includegraphics[width=\linewidth]{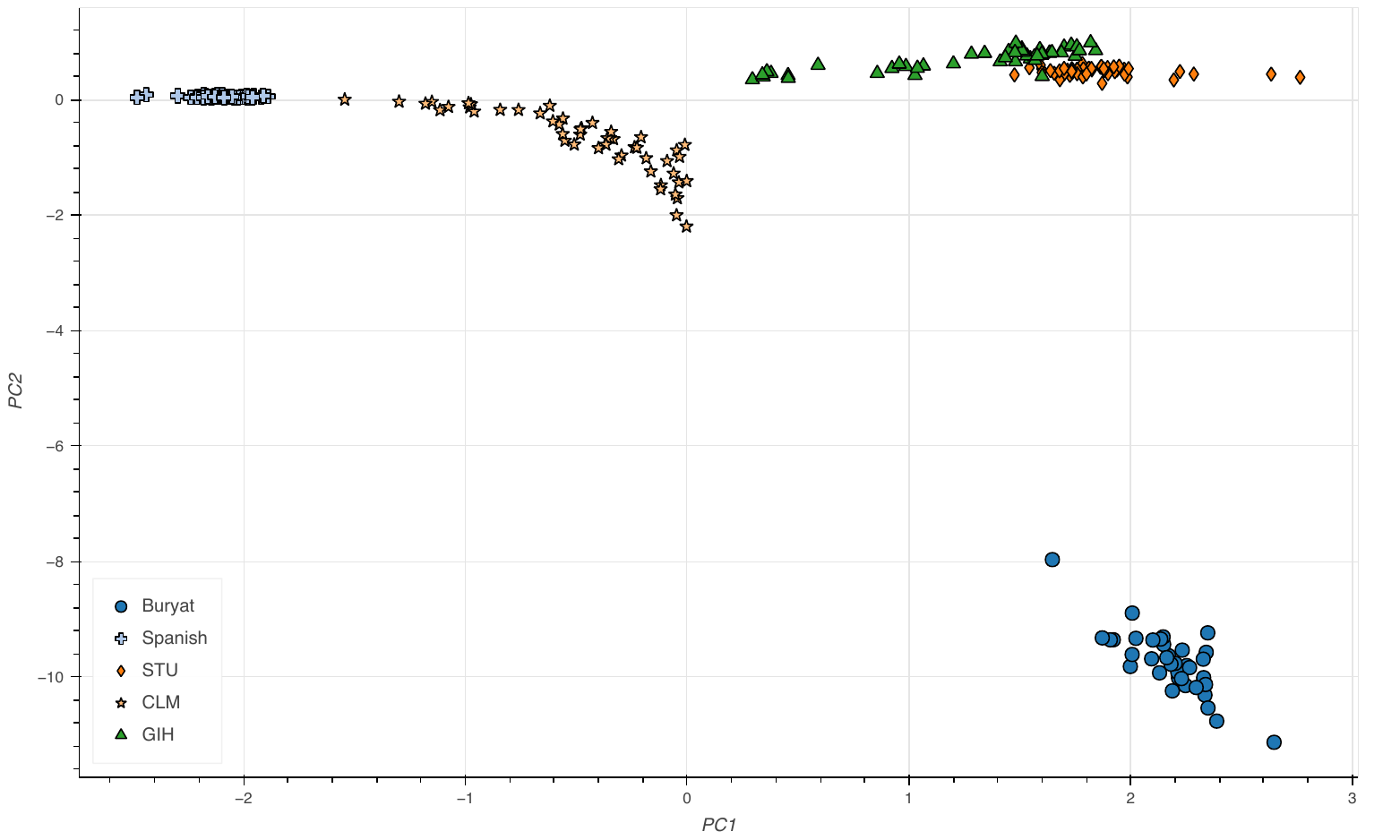}
\subcaption{Proposed Entropy PCA}
\label{fig:pca34}
\end{subfigure}
\vspace{15pt}
\caption{Conventional Principal Component analysis for 5 populations: Buryat, Spanish, Sri Lankan Tamil in the UK (STU), Colombian in Medellín, Colombia (CLM) and Gujarati Indians in Houston, Texas, USA (GIH).
While the gap between CLM and GIH appears rather large in conventional PCA, comparable to the distance between CLM and Buryat, rescaling places CLM substantially closer to GIH, shown in b). }
\label{fig:PCA_ex2}
\end{figure}

\begin{figure}[!ht]
\centering
\begin{subfigure}[t]{\linewidth}
\includegraphics[width=\linewidth]{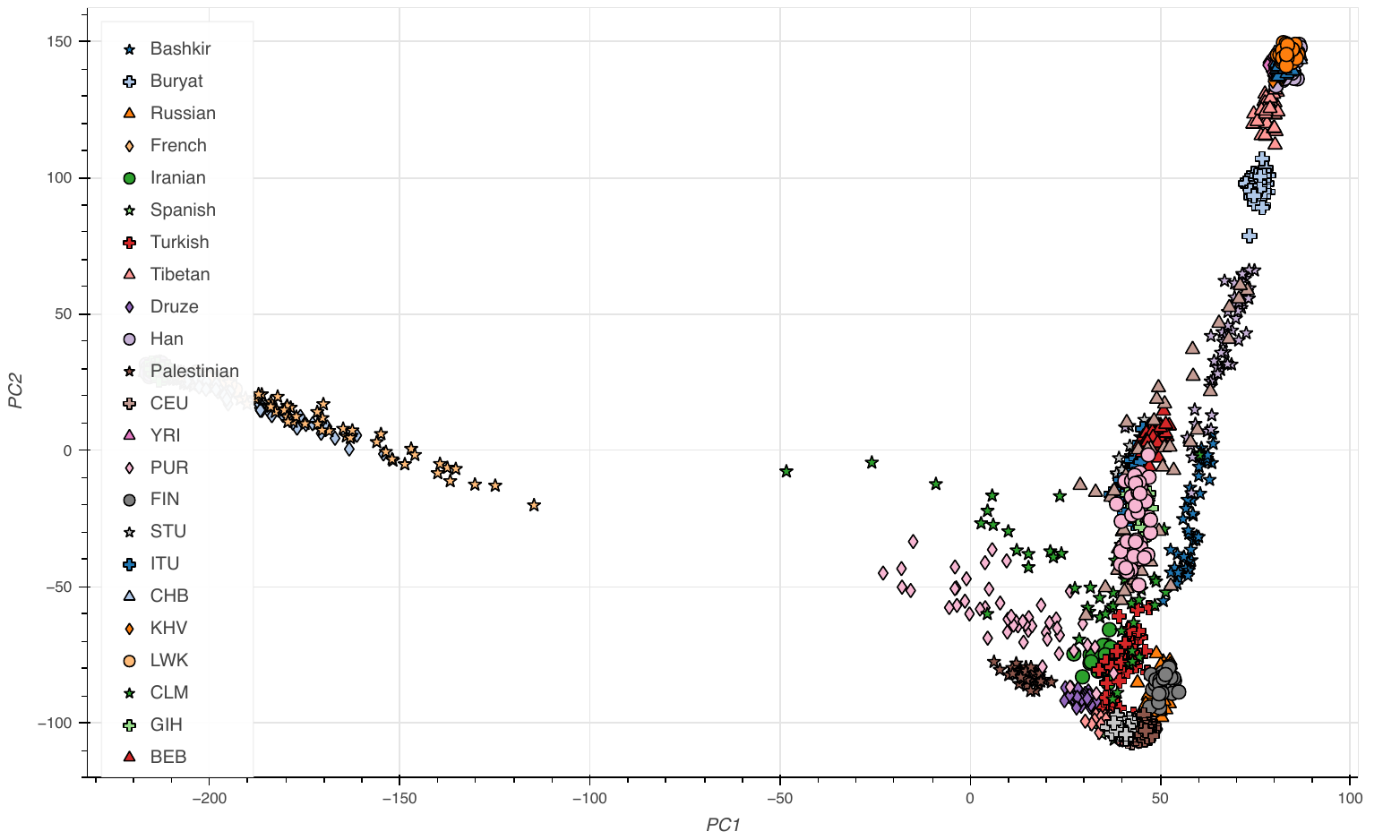}
\subcaption{Correlation-based PCA}
\bigskip
\label{fig:world0}
\end{subfigure}
\hfill
\begin{subfigure}[t]{\linewidth}
\includegraphics[width=\linewidth]{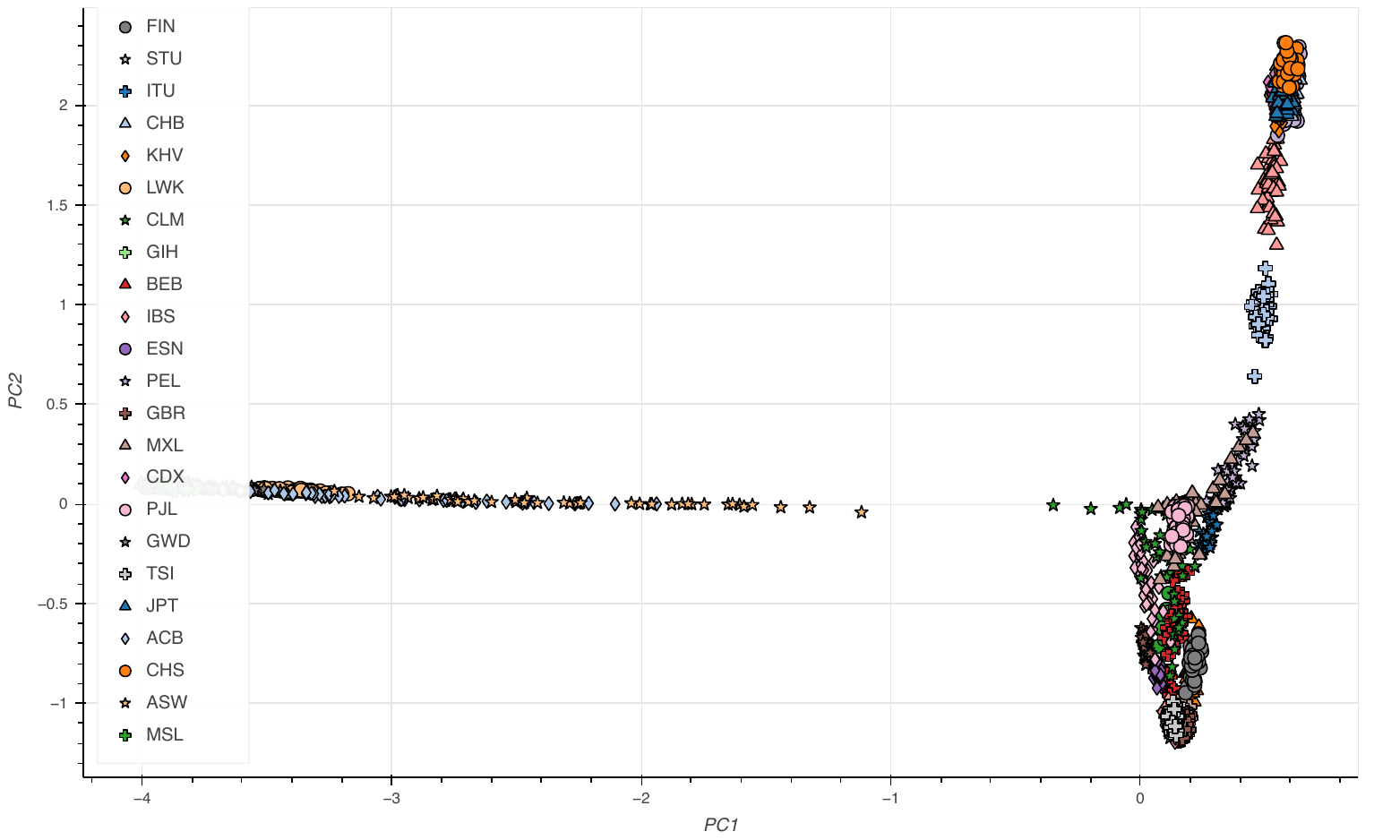}
\subcaption{Proposed Entropy PCA}
\label{fig:world-rescaled}
\end{subfigure}
\vspace{15pt}
\caption{A different world view: the commonly observed triangular PCA shape of world populations undergoes proximity rearrangements using information based rescaling.  Non-African and non-Asian populations are much closer together in b).}
\label{fig:world}
\end{figure}

\subsection{Information and correlation}
It has been shown that experts can be misled by metrics of non-nonlinearity hence the need to "linearize" whatever metric is used. For perceptual constraints compounded by the nonlinearity of the measure; for instance, Soyer et al \cite{soyer2012illusion} showed how great many econometricians, who are well versed statisticians almost always overlook the real inferential and practical implications --all interpretation errors go in one direction, the \textit{fooled by randomness} one (i.e. underestimation of noise). That $70$ pct. of econometricians misinterpreted their own results is quite telling. The corresponding author has documented a version of the effect in \cite{goldstein2007we}: professionals and graduate students failed to realize that they interpreted mean deviation as standard deviation, therefore underestimating volatility, especially under non-Normality. It has been shown in \cite{taleb2020common} that correlation is not additive across subsections of the domain under consideration --even when the variables are Gaussian.

	The only cure --visibly, the evidence is strong tat level of statistical education doesn't help -- is to avoid presenting nonlinear measures and linearizing whatever is presented to the specialist before the scientific implication. 
Entropy methods being additive (unlike correlation) solve the problem. 

Not all branches of research fall for correlation as a relatively uninformational metric. Machine learning loss functions rely on cross-entropy methods \cite{murphy2012machine}. Since DNA is, well, \textit{information}, an information-theoretic metric would be most certainly preferable to what is in current standard use.

Further, since mutual information maps to "how much should one dynamically gamble on $X$ knowing $Y$", its information-theoretic quality is most applicable to genetic distance. Further, in addition to PCA analysis, entropy methods are helpful to properly scale runs of homozygosity (ROH) (that is, contiguous lengths of homozygous genotypes that are present in an individual due to parents transmitting identical haplotypes to their offspring). These are attributes that emerge naturally in phylogenetic situations that PCA has been used to sort out\cite{watterson1975number}.

Other criticisms of PC analysis have been recently made: PC has a spate of weaknesses, much of which are related to sample size mismatch, ability to perform cherry picking, or the insufficiency of representation in two dimensions, as reported in \cite{Elhaik2022}.  However these are standard statistical problems, spreading because of flaws in the applications rather than a fundamental structural problem, and fixable with more rigorous but standard checks. Our approach shows fundamental differences not curable by these conventional checks.

\bigskip
The rest of this discussion is organized as follows. First, we propose a new way to map PCs using mutual information, thanks to the following convenient property. Because PCA vectors for Gaussian variables are orthogonal both for correlation and mutual information, we can apply a simple heuristic for the translation. Second we express the precise mathematics applied to genetics, in matrix form, mapping to our exact implementation on population maps. Finally we show the results as applied to the world population, as well as a subsample of it, with comments on the significant divergences between methods.

\section{PCA under Mutual Information}
We observe that conventional principal component analysis propose distances between groups and variables based on representation on maps built as follows.

Let $(X_1, \ldots, X_n)$ be the original vectors (in $\mathbbm{R}^m$), and $(\pi_1, \ldots,\pi_n)$ the orthogonal principal components ordered by decreasing variance. Two--dimensional principal component representation typically maps  $X_i$ in Cartesian coordinates according to a metric $\mu$ such that the coordinates become 
$$d_i=\left(\mu(X_i, \pi_j) ,\mu(X_i, \pi_{j'})\right)$$
where typically $j' = j+1$. The same logic applies to three dimensions.

The function $\mu(.)$ in common use is expressed by the dot product $<X_i,\pi_j>$ scaled by 
$\frac{1}{m-1}$ , or its decomposition via the scaled correlation 
\begin{equation}
\mu(X_i, \pi_j)=	\rho_{X_i, \pi_j} \sigma_{X_i}\sigma_{\pi_j}
 \end{equation}
and when the $X$ are normalized,
\begin{equation}
\mu(X_i, \pi_j)=	\rho_{X_i, \pi_j} \sqrt{\lambda_j}\label{centraleqPCA}
 \end{equation}
 where $\lambda_j$ is the eigenvalue associated with the principal component $\pi_j$.
 
We will revisit with a matrix notation expressing the suggested transformations as applied to genetic analysis.

\subsection{Mutual Information}
As per the standard definition in the literature\cite{cover2012elements},  $I_{X,Y}$ the mutual information we use between r.v.s $X$ and $Y$:

\begin{equation}
	I_{X,Y}= \int_{\mathcal{D}_X}\int_{\mathcal{D}_Y} f(x,y) \log\left(\frac{f(x,y)}{f(x) f(y)}\right) \; \mathrm{d}x \;\mathrm{d}y
\end{equation}
and of course
$$\log \frac{f(x,y )}{f(x) f(y)}= \log \frac{f(x|y)}{f(x)}=\log \frac{f(y|x) }{f(y)}.$$

In effect, and what is relevant for genetics, mutual information is the Kullback-Leibler divergence between two distributions: the joint distribution $f(x,y)$ and the product  $f(x) f(y)$ evaluated with respect to the joint distribution, \cite{cover2012elements}.

We note some difficulties translating direct frequencies into continuous functions but in our case the problem is solved via the property that for Gaussian distributions, independence implies zero correlation (and vice versa), allowing us to transfer to MI from the pairwise correlation. (We note that common practice consists is smoothing the kernel distribution, then computing the mutual information.)

It is central that, under bivariate normality, the orthogonal principal components satisfy, for $i,j \leq m$
 $$I_{\pi_i, \pi_{j \neq i}}=0.$$


This holds for bivariate normal distributions~\cite{gel1957computation,dynkin_eleven_1959} (though not all distributions in the elliptical class), uncorrelated means independence. Let $\Sigma$ be the covariance matrix for $X, Y \sim \mathcal{N}(\bf{M, \Sigma}$) where $M$ is a vector of means and 
$\bf{\Sigma} =\left(
\begin{array}{cc}
 \sigma _1^2 & \rho  \sigma _1 \sigma _2 \\
 \rho  \sigma _1 \sigma _2 & \sigma _2^2 \\
\end{array}
\right).$  Assume $M=(0,0)$ with no loss of generality. The PDFs are $f(x)=\frac{e^{-\frac{x^2}{2 \sigma _1^2}}}{\sqrt{2 \pi } \sigma _1}$; the joint PDF becomes

$f(x,y) =\frac{\exp \left(-\frac{\sigma _2^2 x^2-2 \rho  \sigma _2 \sigma _1 x y+\sigma _1^2 y^2}{2 \left(1-\rho ^2\right) \sigma _1^2 \sigma _2^2}\right)}{2 \pi \sigma _1 \sigma _2 \sqrt{\left(1-\rho ^2\right) }}.$  So the parametrization $\rho=0$ implies the identity $f(x,y)=f(x) f(y)$, namely that lack of correlation implies independence, hence absence of mutual information between $X$ and $Y$, that is, $I_{X,Y}=0$.


Taking for example other elliptical distributions frequently used in social science, the bivariate Student T or Cauchy, $\rho=0$ does not mean independence \cite{taleb2019statistical}. For instance, for $X, Y\sim$ Multivariate Student T ($\alpha,\rho$), the mutual information $I_\alpha(X,Y)$:

\begin{equation}
	I_\alpha(X,Y)=-\frac{1}{2} \log \left(1-\rho ^2\right)+\lambda _{\alpha }
\end{equation}
where $\lambda _{\alpha }=-\frac{2}{\alpha }+\log (\alpha )+2 \pi  (\alpha +1) \csc (\pi  \alpha )+2 \log \left(B\left(\frac{\alpha }{2},\frac{1}{2}\right)\right)-(\alpha +1) H_{-\frac{\alpha }{2}}+(\alpha +1) H_{-\frac{\alpha }{2}-\frac{1}{2}}-1-\log (2 \pi )$,
where $\csc(.)$ is the cosecant of the argument, $B(.,.)$ is the beta function and $H(.)^{(r)}$ is the harmonic number $H_n^r=\sum _{i=1}^n \frac{1}{i^r}$ with $H_n=H_n^{(1)}$.
We note that for $\lambda _{\alpha } \underset{\alpha \to \infty}{\to} 0 $, the limit that corresponds to the Gaussian case.

 This makes the proposed transformation heuristic more straightforward than alternatives to PCA such as the t-distributed stochastic neighbor embedding (t-SNE) method\footnote{We also note that the (standard) original stochastic neighbor embedding technique does not reflect information-theoretic distances; its aim is to reduce dimensionality.}.
 
We also note Linsker's result \cite{linsker1988self} showing that the conventional PCA provides an information-theoretic optimality so long as the noise is Gaussian.

%

\smallskip

\textbf{Additivity:} We  note that $I_{X,Y}$ is additive across partitions of  $\mathcal{D}_X$ and $\mathcal{D}_Y$, since
$I_{X,Y}= \mathbb{E}\left(\log f(x,y)	\right)-\mathbb{E}\left(\log f(x)	\right) -\mathbb{E}\left(\log f(y)	\right)$. Consider the additivity of measures on subintervals $\int_{A \cup B}  f \, d \mu=\int_{A}  f \, d \mu+\int_{B}  f \, d \mu $.

\subsection{Re-scaling PCA distances using Mutual Information}

 We note that regardless of parametrization of $X$ and $Y$, when the distributions are jointly Gaussian with $\rho_{X,Y}$, $I_{X,Y}=  \frac{1}{2} \log \left(1-\rho ^2\right)$.

 $I_{X,Y}$ the mutual information between r.v.s $X$ and $Y$ and joint PDF $f(.,.)$, because of its additive properties, allows a representation of relative correlations, via the re-scaling function 

 \begin{equation}
 	r_{X,Y}=- \text{sgn}(\rho_{X,Y}) \frac{1}{2} \log \left(1-\rho_{X,Y} ^2\right);
 \end{equation} 	
see Figure \ref{rescaling1} for details. We use the signed correlation to show the direction of the information: MI shows strength of association, not its direction, and for a monotonic linear function, the association is preserved in the negative domain. Hence Eq. \ref{centraleqPCA} can be modified for rescaling (marked as $\mu'$)
 
 \begin{equation}
 	\mu(X_i, \pi_j)'=- \text{sgn}(\rho_{X_i, \pi_j} ) \frac{1}{2} \log \left(1-\rho_{X_i, \pi_j}  ^2\right) \sqrt{\lambda_i},
 	 \end{equation}
as shown in Figs. \ref{transformation}.
%
%
%
%
%
\subsection{In Matrix Notation}
Using matrix notation (mapping to our implementation), we express the problem in the following way.  
Centering and scaling in the correct order yields a matrix suitable for computing correlations.
We start by defining a matrix ${\bf{G}} = (g_{ij})$ features indexed by $i \in \mathbb{Z}_m$ samples, and $j \in \mathbb{Z}_n$. \footnote{These features could be biallelic diploid SNPs coded in $\mathbb{Z}_2$}.  
Define 
\begin{eqnarray}
\mu_i &=& \frac{1}{n-1}\sum_{j\in \mathbb{Z}_n} g_{ij}, \\
\sigma_{ii'} &=& \frac{1}{n-1}\sum_{j \in \mathbb{Z}_n} (g_{ij} - \mu_i)(g_{i'j} - \mu_{i'}), \\
\sigma_i^2 &=& \sigma_{ii} \\
{\bf{Z}} &=& \left(\frac{g_{ij} - \mu_i}{\sigma_i}\right) \\
\rho_{ii'} &=& \frac{\sigma_{ii'}}{\sigma_i \sigma_{i'}}. 
\end{eqnarray}
Then ${\bf{ZZ^T}} = (n-1)\left(\rho_{ii'}\right)$.

Accordingly, ${\bf{ZZ^T}} = \left(\sum_{j\in \mathbb{Z}_n} \frac{(g_{ij} - \mu_i)(g_{i'j} - \mu_{i'})}{\sigma_i \sigma_{i'}}\right) = (n-1)\left(\frac{\sigma_{ii'}}{\sigma_i \sigma_{i'}}\right) =  (n-1)\left(\rho_{ii'}\right)$.
Therefore the correlation matrix $\bf{C}$ may be represented \footnote{The transpose aims at ascertaining that in some software programs such as Mathematica, the eigenvectors are presented as the columns of the matrix.} by \footnote{The centering by rows  for genotypic analysis differs from Patterson\cite{patterson_population_2006}, but conforms with Price et al.\cite{price_principal_2006}; the ``smartpca'' app computes the appropriate correlations with ``altnormstyle: NO''.}

\begin{equation}
    {\bf{C}} = \left(\rho_{ii'}\right) = \frac{1}{n-1}{\bf{ZZ^T}} = \mbox{cov}({\bf{Z}} ,{\bf{Z^T}}).
\end{equation}

 $\bf{C}$ is symmetric and positive definite.   
Since, for any vector $\bf{w}$, the expression ${\bf{w^T C w}} = \frac{1}{n-1}({\bf{Z^T w}})^T({\bf{Z^T w}}) \ge 0$, it follows $\bf{C}$ is positive definite.  Also, ${\bf{C}^T} = \frac{1}{n-1}({\bf{ZZ^T}})^{\bf{T}} = \frac{1}{n-1}{\bf{ZZ^T}}={\bf{C}}$, and so is symmetric.

The diagonalization of $\bf{C}$ provides a decomposition of the feature vectors into an orthogonal set that spans the subspace containing the samples.
The $\bf{{U^T Z}}$ rows are orthogonal, and the covariance diagonal.    

Given that $\bf{C}$ is positive definite and symmetric,  ${\bf{C}}$ is diagonalized by an orthonormal matrix ${\bf{U}}$ of the normalized orthogonal eigenvectors to yield a diagonal matrix $\bf{D}$, so that ${\bf{CU}} = {\bf{UD}}$. ${\bf{S}}^2 = (n-1){\bf{D}}$ is in common usage so that $({\bf{Z}}{\bf{Z^T}}){\bf{U}} = {\bf{US^2}}$. Therefore ${\bf{D}} = {\bf{U^T C U}} =  \mbox{cov}(({\bf {U^T Z}}), ({\bf {U^T Z}})^{\bf{T}}) = \frac{1}{n-1}{\bf{U^T Z Z^T U}}$.  Since $\bf{D}$ is diagonal, the $\bf{{U^T Z}}$ rows are orthogonal, and the covariance $\bf{D}$ in that basis is diagonal.     

We can identify the $n$ columns, $m$ rows, matrix of $n$ feature-wise orthogonal principal components $\pi_i$ as:

    \begin{equation}
    \bf{P}=\bf{U^T} \bf{Z} 
    \end{equation}

Note that, since the covariances of the $\bf{P}$ are $\mbox{cov}(\bf{P}, \bf{P^T}) = \bf{D}$ is diagonal, the rows are orthogonal, as noted previously.  The matrix

\begin{equation}
{\bf{V}} = (n-1)^{-1/2} {\bf{D^{-1/2}} \bf{P}} = {\bf{S}}^{-1} {\bf{U^T Z}}
\end{equation}

is normalized so that $\bf{V} \bf{V^T} = I$.  $\bf{V}$ is half-orthornormal; the transposes are not: $\bf{V^T} \bf{V} \ne I$.  The reason for this is that the number of individual vectors of SNPs for the individuals in $\bf{Z}$ does not span the space of SNP vectors since $m \ll n$.  These are the familiar matrices in the singular value decomposition

\begin{equation}
\bf{Z} = \bf{U}\bf{S}\bf{V^T}.    
\end{equation}

This decomposition also shows that the vectors in $\bf{V^T}$ represent an orthogonal basis in which $\bf{Z}$ can be represented, and so spans the subspace spanned by $\bf{Z}$.  

Also $\mbox{cov}(\bf{S},\bf{S^T}) = \bf{U^T}\mbox{cov}(\bf{Z}, \bf{Z^T})\bf{U}$ will be useful.

We define the correlation matrix

\begin{equation}
\bf{M} = \mbox{cor}(\bf{Z}, \bf{P^T})
\end{equation}    

Then
\begin{equation}
\bf{M} = \bf{U}.
\end{equation}
$\bf{M} = [\mbox{cov}(\bf{Z},\bf{Z^T})]^{-1/2} \mbox{cov}(\bf{Z},\bf{P^T}) [\mbox{cov}(P,P^T)]^{-1/2} $.  Noting that $\mbox{cov}(\bf{Z},\bf{Z^T})
= \frac{1}{n-1}\bf{U^T}\bf{S^2}\bf{U}$, $\mbox{cov}({\bf{P}},{\bf{P^T}}) = \frac{1}{n-1}{\bf{S^2}}$, and $\mbox{cov}({\bf{Z}}, {\bf{P^T}}) = \frac{1}{n-1}{\bf{ZZ^TU}}=\frac{1}{n-1}{\bf{US^2}}$, then $\bf{M} = \bf{U}\bf{S}^{-1}\bf{U^T} {\bf{U}}{\bf{S^2}} {\bf{S}^{-1}} = U$.
This is therefore the standard principal component matrix that we expect, {\it and}, since this is a correlation, this may be re-scaled as a mutual information.
The information re-scaled version $\bf{M'}$ becomes

$$
	\bf{M'}=R(\bf{M}) = R(\bf{U}).
$$

%
%
%
%
%
%
%

\section{Discussion and Application to Genetic Distance}

We investigate the visualization of genetic distances in world populations. To this end we select populations from the AADR Human Origin dataset version 54.0, which in turn comprises samples from the 1000 genome project\cite{1000_genomes_project_consortium_global_2015}. The following populations were selected: Peruvian from Lima (PEL, 68 samples), Punjabi in Lahore, Pakistan (PJL, 96), Bashkir (53), Central European in Utah (CEU, 99), Chinese from Xishuangbanna (CDX, 93) and Han Chinese (45).
The microarray dataset contains 597,573 typed loci.
Total genotyping rate is 0.999255.


Figure \ref{fig:world} provides a new perspective on the common PCA plot of world populations derived from the 1000 Genome Project. 
 While the Asian and African clines still determine the overall structure (though nearer to the axes), the remaining world populations are closer together. For example, for Puerto Ricans (PUR) and Colombians (CLM) conventional PCA spreads them along the beginning of the African cline, whereas rescaling shows them in the vicinity of other Latin American populations (Mexicans and Peruvians).

Iranians, Turkish, Palestinian, Druze, French, Iberian (IBS), British (GBR), Russian, Finnish, Puerto Rican, and the majority of Colombians all form a much tighter cluster in the rescaled PCA, indicating that these populations are not as far from each other as the conventional PCA suggests.

%
%

To conclude, we showed how, under some conditions satisfied in population genetics, to efficiently and effectively convert a principal components based map to one representing information-based distance. There are more than $200,000$ published results that may be affected by this simple change of metric \cite{Elhaik2022}, with results that would need to be reevaluated.


\begin{appendix}

\setcounter{section}{0}
\renewcommand{\thesection}{\Alph{section}}
\section{Supplementary Material}
The following populations were taken from the AADR set (exact population labels as in AADR are used):
\begin{table}[h!]
\begin{tabular}{ll|ll}
   Samples & Populations          & Samples &     Populations \\

   92 &      ACB.DG &     102 &          ITU.DG \\
   55 &      ASW.DG &      38 &      Iranian.HO \\
   85 &      BEB.DG &     104 &          JPT.DG \\
   53 &  Bashkir.HO &      95 &          KHV.DG \\
   37 &   Buryat.HO &      99 &          LWK.DG \\
   93 &      CDX.DG &      85 &          MSL.DG \\
   99 &      CEU.DG &      62 &          MXL.DG \\
  103 &      CHB.DG &      68 &          PEL.DG \\
  103 &      CHS.DG &      96 &          PJL.DG \\
   94 &      CLM.DG &      99 &          PUR.DG \\
   41 &    Druze.DG &      38 &  Palestinian.DG \\
   99 &      ESN.DG &      71 &      Russian.HO \\
   97 &      FIN.DG &      98 &          STU.DG \\
   61 &   French.HO &     172 &      Spanish.HO \\
   90 &      GBR.DG &     107 &          TSI.DG \\
  102 &      GIH.DG &      97 &      Tibetan.HO \\
  112 &      GWD.DG &      50 &      Turkish.HO \\
   45 &      Han.DG &     101 &          YRI.DG \\
  103 &      IBS.DG &        &                \\

\end{tabular}
\end{table}
\end{appendix}

\begin{thebibliography}{999}
\bibitem{taleb2019statistical}
N.~N. Taleb, \emph{The Statistical Consequences of Fat Tails}.\hskip 1em plus
  0.5em minus 0.4em\relax STEM Academic Press, 2020.

\bibitem{soyer2012illusion}
E.~Soyer and R.~M. Hogarth, ``The illusion of predictability: How regression
  statistics mislead experts,'' \emph{International Journal of Forecasting},
  vol.~28, no.~3, pp. 695--711, 2012.

\bibitem{goldstein2007we}
D.~Goldstein and N.N.~Taleb, ``We don't quite know what we are talking about when
  we talk about volatility,'' \emph{Journal of Portfolio Management}, vol.~33,
  no.~4, 2007.

\bibitem{murphy2012machine}
K.~P. Murphy, \emph{Machine learning: a probabilistic perspective}.\hskip 1em
  plus 0.5em minus 0.4em\relax MIT press, 2012.

\bibitem{Elhaik2022}
E.~Elhaik, ``Principal Component Analyses (PCA)-based findings in population genetic studies are highly biased and must be reevaluated,'' \emph{Scientific Reports}, vol.~12, no.~1, 14683, 2022.


\bibitem{taleb2020common}
N. N. ~Taleb, ``Common misapplications and misinterpretations of correlation in
  social science,'' \emph{Preprint, Tandon School of Engineering, New York
  University}, 2020.

\bibitem{cover2012elements}
T.~M. Cover and J.~A. Thomas, \emph{Elements of information theory}.\hskip 1em
  plus 0.5em minus 0.4em\relax John Wiley \& Sons, 2012.

\bibitem{gel1957computation}
I.~M. Gel'fand and A.~M. Yaglom, ``Computation of the amount of information
  about a stochastic function contained in another such function,''
  \emph{Uspekhi Matematicheskikh Nauk}, vol.~12, no.~1, pp. 3--52, 1957.

\bibitem{dynkin_eleven_1959}
A.~Gel'fand, I.M.;~Yaglom, ``{Calculation of amount of
  information about a random function contained in another such function},'' in
  \emph{ {Eleven {Papers} on {Analysis}, {Probability}
  and {Topology}}}, American Mathematical Society, Dec. 1959, vol.~12, iSSN:
  0065-9290, 2472-3193. 

\bibitem{linsker1988self}
R.~Linsker, ``Self-organization in a perceptual network,'' \emph{Computer},
  vol.~21, no.~3, pp. 105--117, 1988.

\bibitem{patterson_population_2006}
N.~Patterson, A.~L. Price, and D.~Reich, `` {Population
  {Structure} and {Eigenanalysis}},'' \emph{ {PLOS
  Genetics}}, vol.~2, no.~12, p. e190, Dec. 2006, publisher: Public Library of
  Science. 
  
  \bibitem{price_principal_2006}
A.~L. Price, N.~J. Patterson, R.~M. Plenge, M.~E. Weinblatt, N.~A. Shadick, and
  D.~Reich, `` {Principal components analysis corrects
  for stratification in genome-wide association studies},''
  \emph{ {Nature Genetics}}, vol.~38, no.~8, pp.
  904--909, Aug. 2006, number: 8 x
  
  \bibitem{1000_genomes_project_consortium_global_2015}
1000 Genomes Project Consortium, ``A global reference for human genetic variation,'' \emph{Nature}, vol.~526, no.~7571, p. 68, 2015, publisher: Nature Publishing Group.


\bibitem{watterson1975number}
GA,~Watterson, ``On the number of segregating sites in genetical models without recombination,``\emph{Theoretical population biology},vol.~7, no. ~12, pp 256--276.





\end{thebibliography}
\end{document}